\documentclass[aps, preprint, showpacs, superscriptaddress]{revtex4}
\usepackage{graphicx}
\usepackage{amsmath}
\usepackage{bm}
\begin{document}
\title
%---------------------------------------------------------
{Evidence for hyperconductivity and thermal superconductivity}
%---------------------------------------------------------

\author{V.A. Vdovenkov}

\affiliation{Moscow State Institute of Radioengeneering, Electronics and Automation (technical university),
Vernadsky ave. 78, 117454 Moscow, Russia}

\begin{abstract}
Physical explanation of hyperconductivity and thermal superconductivity existence is done in given article on the basis of  inherent atomic nuclei oscillations in atoms of materials which are connected with electrons and phonons and in accordance with the well known Bardeen-Cooper-Schrieffer superconductivity theory.

It is shown that hyperconductivity is the self-supporting, independent physical phenomenon which is caused by oscillations of atomic nuclei in atoms of materials and the minimal temperature of its existence does not reach absolute zero temperature. Hyperconductivity represents the typical dynamic condition of a material with zero electrical and zero thermal resistances.
\end{abstract}
% 61.82.Fk  Semiconductors
% 63.20.Mt  Phonon-defect interactions
% 65.00.00  Thermal properties of condensed matter
% 66.00.00  Transport properties of condensed matter (nonelectronic)
% 74.25.-q  General properties; correlations between physical properties in normal
%             and superconducting states
% 74.72.-h  High-Tc compounds

\maketitle

%\keyword{Hyperconductivity, thermal superconductivity, nuclei oscillations.}

\section{Introduction}
According to experiments the electrical and thermal resistances of semiconductors sharply decreases at heating above characteristic temperatures of transition to hyperconductance state ($T_h$) if they contain the so-called electron-vibrational centers (EVC). Such containing EVC semiconductors are hyperconductors.

EVC represent such local centers, whose equilibrium positions or frequencies of elastic oscillations in a material depend on their electronic condition. In transitions of electrons or holes  on EVC energy levels inevitably participate elastic oscillations of a material (phonons and nuclei oscillations in atoms of EVC). In this connection such transitions are electron-vibrational transitions. EVC and connected to them the electron-vibrational transitions  essentially change physical properties of materials. Apparently, the first successful researches EVC have been executed in connection with studying of the so-called color centers in dielectric materials \cite{pekar1951, pekar1952, pekar1953, huang1950}. It  turned out later, that EVC in semiconductors give rise to hyperconductivity at temperatures above $T_h$.

Values $T_h$ may be calculated by taking into account the material parameters, EVC concentration (N) and values of the connection constants between electrons and phonons (S) \cite{vdov2000,  vdov2001, vdo2002, vdov2002}. So, that experimental values $T_h $, in dependence from average value of a material atomic number  ($Z _ {mid} $), occupy band with the well known width about $300 K $. Borders of this band coincide with the calculated maximal and minimal values $T_h$ according to maximal and minimal EVC concentrations in materials. Increase $Z_{mid}$ corresponds to reduction $T_h$ and on the contrary, reduction $Z_{mid}$ corresponds to increase $T_h$. Heating of such material is higher $T_h$ causes sharp and unguided reduction of its electric resistance. Resistance remains lowered and even zero in the temperature range including near room temperatures and higher temperatures, beginning from $T_h$ up to $800 K$ and is higher, presumably above of a material melting and in liquid material. Two peaks of thermopower (Seebeck effect) with different polarity precede the sharp reduction of electric resistance, whereupon thermal and electrical resistances diminish practically up to zero. It show the phenomena of hyperconductivity and thermal superconductivity. These (and some others) physical properties are characteristic only for EVC containing semiconductors and undoubtedly are caused by the EVC.

All in all, hyperconductivity is identical to superconductivity, but differs from superconductivity by the physical mechanism of the new phase formation. Hyperconductivity arises at heating of semiconductors containing EVC above $T_h$, but superconductivity arises at cooling of superconductors below temperature of superconducting transition ($T_c$). Existence of hyperconductivity sometimes excites some doubts and consequently demands for explanation, an establishing of hyperconductivity connection with known physical theories and representations. This task appears simple in connection with discovery of inherent (I-) oscillations of atomic nuclei in atoms of materials, and also due to known connection of EVC I-oscillations with hyperconductivity.

The main purpose of given article is to prove physically the opportunity of hyperconductivity existence as independent physical phenomenon and establish the quantitative and qualitative connection of hyperconductivity with superconductivity.

\section{Hyperconductivity comparison with superconductivity}
The superconductivity, discovered by Kamerling Onnes in 1911,  traditionally is considered as the low-temperature phenomenon. One of the important parameters of every superconductor material is the characteristic temperature of its superconducting transition $T_c$. Cooling down of a superconductor up to temperature $T_c$ gives rise to sharp decrease practically up to zero of its electric resistance which remains the same at the further cooling.  By present time, many superconducting materials with various values $T_c$ are discovered experimentally, empirically. High-temperature superconductors for which value $T_c$ exceeds $20 K$ are picked out among them. Perovskite-like metallic oxide have value $T_c$ reaching $134 K$. Superconductors with $T_ñ \cong 200 K$ are known \cite{gulian2005}. Searching for superconductors with temperatures of superconducting transitions in room temperatures area and higher is continuing. An opportunity of superconductivity existence at room temperatures and at higher temperatures has not been proved, but it also has not been disproved.

The superconductivity theories describe properties of superconducting conditions, but, unfortunately, do not predict superconducting properties of materials. Therefore known superconductors are found out empirically, without the  theoretical forecast, practically "blindly". Meanwhile the principal opportunity of superconductivity existence even at very high temperatures is in known theories of superconductivity, in particular in (BCS) theory of J. Bardeen, L. N. Cooper, J. R. Schrieffer  \cite{bard1957}.  Such opportunity was known long ago, tens years ago, and consists in formation of a superconducting condition with participation of phonons of materials with big energies. Discovery of inherent oscillations of atomic nuclei in atoms of materials (having the big energy quantums, exceeding energy of optical and acoustic phonons in some times \cite{vdo2002, vdov2002, vdov2003} has allowed us to carry out this opportunity practically. Using of inherent oscillations allows to bring about hyperconductivity and typical for it zero electrical resistance and zero thermal resistance even at very high temperatures.

Theories explain superconductivity by effective connection between electrons and elastic oscillations of materials that are in the consent  with results of optical, microwave, ultrasonic and other researches, and also was confirmed by isotopic effect. In particular the BCS theory allows to define the basic physical parameters of superconductors and to calculate characteristic temperatures of superconducting transition
\begin{equation}
\label{eq1}
T_c = 1.13T_D exp(-[V^*N(F)]^{-1}),
\end{equation}
where $T_D = \hbar\omega_D/k$  - Debye temperature of elastic oscillation of a material which is connected with electrons (or holes), $\hbar\omega_D$ - quantum and $\omega_D$ - Debye frequency of the oscillation,  $\hbar$ - Planck constant, $k$ - Boltsman constant, $V^*$  - energy of connection with electron of elastic material oscillation, $N(F)$  - density of electronic conditions at Fermi energy, $F$ - Fermi energy. For example, energy livel $F$ is located in the middle of forbidden energy gape of metallic superconductors (2$\Delta$) where $N(F)$ and $V^*$ are small. It is usualy $T_D < 200 K$. Hence $V^*N(F) << 1$ and  $T_c$ does not exceed 20 K. On the contrary,  in semiconductors containing EVC  (i.e. in hyperconductors) $T_D > 200 K$, $V^*N(F) > 1$  and calculated values $T_c$ exceed the melting temperatures of materials ($T_{melt}$).

Really, ones believes as usual, that Eq.~\eqref{eq1} is true only for low temperature superconductivity. This opinion cannot be considered as final decision because in  materials rather energetic elastic oscillations are possible, and increasing of their connection with electrons increase the value $T_c$ repeatedly. It is evident from Eq.~\eqref{eq1} that large value $T_c$ may be achieved when elastic oscillations of a material with high Debye frequencies $\omega_D$ (with large Debye temperatures $T_D$) are connected with electrons strongly, when energy $V^*$ and density of electronic conditions $N(F)$ are large enough.

These electrons, connected with elastic oscillations of a material, are superconducting electrons. They provide superconducting properties of materials. Such electrons, generally speaking, are not free, but in BCS theory they submit to Fermi - Dirac  distribution as though they form ideal gas of superconducting electrons. The similar result turns out in the Shockley-Reed theory of electrons and holes recombination where distribution function of electrons, localized on deep energy levels in the forbidden energy gape of semiconductor, is calculated and also coincides with Fermi-Dirac function.

Elastic I-oscillations of atomic nuclei inside atoms or ions can exist in materials in addition to acoustic and optical oscillations. $\alpha$-, $\beta$- and $\gamma$- types of the oscillations and waves of such oscillations are known \cite{vdov2000, vdov2002, vdo2002, vdov2003}. Energies of I-oscillations  may be described by the formula of linear quantum harmonious oscillator
\begin{equation}
\label{eq2}
E(\nu)=\hbar\omega_Z(\nu+1/2),
\end{equation}
where $\hbar\omega_Z$ - energy quantum of I-oscillations $\alpha$-, $\beta$- or $\gamma$-type in atom with number $Z$, oscillatory quantum number $\nu$ can accept values 0, 1, 2, ...  Smaller among the specified $\hbar\omega_Z$ concern to oxygen atom ($Z = 8$), namely: $\hbar\omega_8 \cong 0.22 eV$  and $T_D \cong 2640K$ for $\nu = 0$.

One can see from Eq.~\eqref{eq1} that due to high-energies $E(\nu)$ and high  temperatures $T_D$ the calculated temperatures of superconducting transition $T_c$ exceed $1490 K$ for $\nu = 0$ and can be considerably higher when $\nu > 1$. However, it is necessary to excite I-oscillations of atomic nuclei in a material and to support enough strong  connection of these oscillations with electrons for achievement of such high  temperatures of superconducting transitions. Both these conditions are met in semiconductors containing the so-called electron- vibrational centers - EVC. EVC create conditions for realization of semiconductors hyperconductivity, that is superconductivity at high temperatures and at very high temperatures. Hyperconductivity is caused by I-oscillations of EVC.

Typical EVC in monocrystal silicon are associations of impurity oxygen atoms with vacancies ($A-centers$) \cite{bems1959, stone1975}. A-centers usually introduce in a material by irradiating silicon by fast electrons. However, EVC is possible to create in materials by other ways.
Conditions for occurrence and existence of semiconductors hyperconductivity are satisfied automatically at temperatures $T > T_h$ (when semiconductors contain EVC). Thermal generation of charge carriers becomes sufficient for excitation of necessary density of I-oscillations at the expense of recombination energy of electrons and holes on EVC.

Experiments have shown, that EVC concentration (N), necessary for hyperconductivity, sit between $N _ {min} = 2\cdot10 ^ {12} sm ^ {-3} $ and $N _ {max} = 3\cdot10 ^ {17} sm ^ {-3} $, and the values $T_h $ sit in the temperature interval calculated for whatever semiconductor. For example, in monocrystal silicon with nuclear number Z = 14 and width of forbidden gape $E_g \cong 1.1 eV$~~$T_h$ may accept values between 220 K and 530 K. In semiconductor CdHgTe with average value of nuclear number $Z_{mid} = 90$ and $E_g \cong 0.1 eV$ $T_h$  may accept values between 60K and 350K.

Thus, superconductors and hyperconductors have the identical features as both those and others can be in a condition with zero electric resistance. However they have distinctions also.
As against superconductors, hyperconductors are characterized by two characteristic temperatures. Namely: by the transition temperature to hyperconductivity $T_h$ and by extremely high temperature of  superconductivity transition $T_c$. Hyperconductors pass into condition with zero resistance at heating a material above character temperatures $T_h $ and have zero electric and thermal resistances at temperatures between $T_h $ and $T_c $.

According to experiments the temperature dependences of electric resistance of hyperconductors are described by several activation energies. In the region of temperatures is lower than $T_D = 230 K$ these energies coincide with energies of acoustical and optical phonons in material. In the region of temperatures is higher than $T_D$  the activation energies coincide with energies, calculated under the Eq.~\eqref{eq2}.

The voltage-current (VC) characteristics of hyperconductors at temperatures is lower $T_h$ submit to the Ohm law  in weak electric fields with intensity $E \leq 30 V/cm$. At greater intensity $E$ the VC characteristics are nonlinear, have the step structure and aspire to saturation up to electric breakdown of a material  (at $E > 10^4 V/cm$). Intensity of E at electric breakdown of hyperconductors nonmonotonic reduces depending on an induction of a magnetic field $B \leq 2$ Òåñëà, directed along a normal to a current. In magnetic fields with $B > 1$ Tesla the VC characteristics are described by empirical formula $I = I_o E^{\eta}$ where $I_o$ does not depend on $E$ and $\eta$ consistently accepts meanings 1.0, 1.25, 2.5, 5.0, 10.0 at E increasing.

It is impossible to explain the experimental temperature dependences of resistance and  singularity of VC characteristics of hyperconductors on basis of Fermi - Dirac distribution  for electrons and holes, but there is coordination wtith Gauss distribution.

If to take into account that $T_c$ for hyperconductor exceeds melting temperature of a material ($T_{melt}$) then formally it is possible to speak about superconducting transition at temperature $T_c$. However, it is necessary to take in to account that zero resistance of liquid hyperconductor is possible as frequencies of inherent oscillations of nuclei at material melting do not change. Really, frequencies and amplitudes of inherent nuclei oscillations  in atoms or ions do not depend on  presence or absence of several valence electrons. The constant $S$ do not diminish sufficiently after melting of material  also. Phonons with high energies, which necessary for hyperconductivity, disappear at temperatures below $T_h$ and then hyperconductivity collapse occur. Thus hyperconductivity is superconductivity, for which the least existence temperature $T_h$ does not reach absolute zero.

The given facts have removed our representation about superconductivity as about the extremely low-temperature  phenomenon and have practically removed temperature restrictions for realization of zero electric and thermal resistances of condensed materials. The new physical opportunities became a reality due to using of I-oscillations of atomic nuclei for realization of hyperconductivity. In a result it is possible to tell, that hyperconductivity represents a new physical condition of  the EVC containing semiconductor materials with zero electric (and thermal) resistance. Hyperconductivity represent superconductivity at temperatures from the neighborhood of room temperature ($T_h$) to higher temperatures.
\section{Energy diagram of hyperconductor}
\begin{figure}[htbp]
\vspace*{-0.5cm}
%\mbox{\epsfxsize=3.3in \epsfbox{Fig1.eps}}
\includegraphics[width=3.3in]{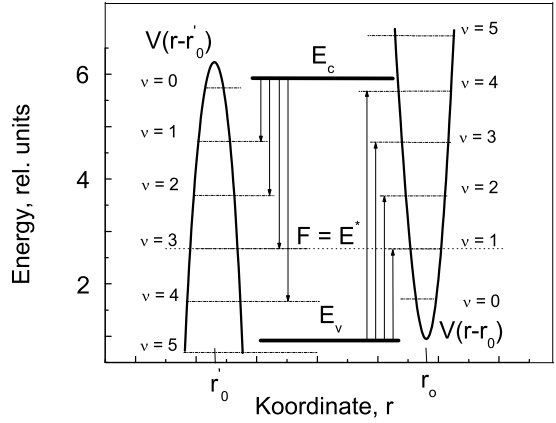}
\vspace*{-0.5cm}
\caption{Energy diagram of a hyperconductor.}
\label{fig1}
\end{figure}

Electron-vibrational energy levels of EVC are described by the Eq.~\eqref{eq2}. They appear in semiconductors as the so-called deep energy levels in the forbidden gape of the semiconductor. According to the data about recombination of electrons and holes on EVC, some of EVC electron-vibrational levels really belong to  forbidden gape of the semiconductor, as shown on Fig.~\ref{fig1}. Energy scheme of a semiconductor is represented in the center of Fig.~\ref{fig1} where $E_c$ and $E_v$ designate energy of bottom of a conductivity zone and a top of a valence zone, F - Fermi level. The considered EVC are located in volume of the semiconductor, in a points with coordinates $r_0$ and $r^{'}_{0}$.
Parabolic potentials, which keep atomic nuclei close by centers of electronic envelops in EVC, are represented by curves $V(r-r_{0})$ in the right part  and $V(r - r^{'}_{0})$ in the left part of Fig.~\ref{fig1}. Electronic transitions from conductivity zone to oscillatory EVC levels  with $\nu > 0$ are shown on Fig.~\ref{fig1} by vertical arrows, which are directed downwards from $E_c$.
Holes transitions to electron-vibrational EVC levels also can be submitted by the vertical arrows directed upwards from $E_v$. In this case the branches of parabola $V(r-r^{'}_{0})$ need to be turned upwards, as shown on the left part of Fig.~\ref{fig1}. Such disposition of potential curves corresponds to excitation of EVC I-oscillations at the expense of energy of electronic (or holes) transitions. These transitions on EVC mainly occur with emission or absorption several phonons, raise I-oscillations of atomic nuclei in atoms of EVC and consequently they are electron-vibrational transitions. Electron-vibrational process on EVC can be presented as consecutive, periodic alternation of electron-vibrational transition from conductivity zone (or from a valent zone) on EVC and the subsequent emission of electron (hole) from EVC at the expense of I-oscillations energy of atomic nuclei and phonons. Each atom, in which nucleus carry out free or compelled I-oscillations, is possible to consider as I-oscillator.  Energies of the oscillations are described by the eq.~\eqref{eq2}.

\begin{figure}[htbp]
\vspace{+0.2cm}
%\mbox{\epsfxsize=3.3in \epsfbox{Fig2.eps}}
\includegraphics[width=3.3in]{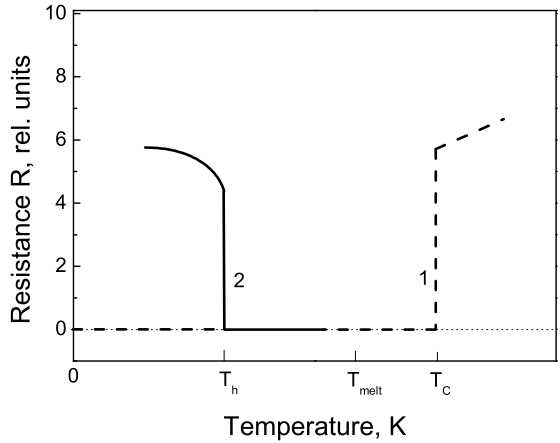}
\vspace{-0.3cm}
\caption{Temperature dependencies of electrical resistance of hyperconductor R. Dashed curve 1 is predicted by BKS theory. Continuous curve 2 is  experimental, typical for  hyperconductor.}
\label{fig2}
\end{figure}

EVC have big capture cross section of electrons (holes), due to strong electron-phonon connection. Electrons and holes become localized on specified electron-vibrational levels of EVC. One of the levels with quantum number $\nu = \nu^*$ dominates in recombination processes. In result, electric charges collects on the given energy  level,  Fermi level $F$ is being fixed on this level $F = E(\nu^*) = E^*$. Density of electron-vibrational conditions on Fermi level $N(F) = N^*\delta(E^*-F)$, where $N^*$  - density of conditions with energy $E^*$, $\delta(E^*-F)$ - delta function.  It is obvious, that $N^*$ exceeds the average density of conditions $N/\hbar\omega_Z$  , where $N$ - concentration of EVC, and it is visible from Fig.~\ref{fig1}, that $V^* = (E_c - E^*) \geq \hbar\omega_Z$. Even if on each center is located one electron (on the average) then for unit volume of a material ($1 sm^3$) value  $V^*N(F) > 2\cdot10^{12}$ because, in accordance with experimental data, $N_{min} \cong 10^{12} cm^{-3}$. Therefore exponent in Eq.~\eqref{eq1} turns into unit, and the calculated temperature $T_c \cong 1.13T_D$. For example, for silicon, containing A-centers, $T_c > 2900 K$ at $\nu \geq 1$. Temperature dependence of electric resistance (R) of a hyperconductor, i.e. containing EVC  semiconductor, calculated on basis of BKS theory, is shown qualitatively on Fig.~\ref{fig2} by dashed line 1.  Value $R$ coincides with $R = 0$ on a site from $T = T_c$  to  $T = T_h$. Typical calculated and experimental temperature dependences of a hyperconductor resistance are shown on Fig.~\ref{fig2}.
One can see from Fig.~\ref{fig2} that resistance of a hyperconductor becomes zero at temperatures higher than temperature $T_h$, but at $T < T_h$ $R$ has nonzero value. Temperature $T_h$ can be determined, using parameters of a material. Really, according to the theory of electron-vibrational transitions \cite{pekar1951, pekar1952, pekar1953, huang1950} in such transition on EVC participate $S$ phonons on the average. Speed of thermal generation of electrons in a stationary condition in a material should be equal to their recombination speed. Next expression connecting concentration N of EVC centers, temperature hyprconductivity transition $T = T_h$ and a constant of electron-phonon connection $S$ follow from this condition:
\begin{equation}
\label{eq3}
            n_i = N_cN_v exp(-E_g/2kT) = N/{S exp(E_{\nu}/kT -1)},
\end{equation}
where $n_i$ - inherent concentration of charge carriers, $N_c$ and $N_v$ - effective values of states for electrons in conductivity band and for holes in valence band of the semiconductor, $E_{\nu}$  -  energy of EVC  inherent oscillations calculated in accordance with Eq.~\eqref{eq2}.

It is possible to estimate the value of $S$ ( constant of electron-phonon connection), incorporated in Eq.~\eqref{eq3}, by considering the energy diagram represented on Fig. 1. Really, in the forbidden energy gape of a semiconductor are no more than $S^{'} = (E_c - E_v)$/$\hbar\omega_Z$   of EVC energy levels. It is obvious, that elastic oscillations of EVC with $\nu >> S{'}$ are improbable, because their general energy exceeds width of the forbidden gape $E_{g} = E_{c} - E_{v}$ and EVC oscillation energy is mainly spent on generation of electron-hole pairs. Therefore in electron-vibrational transitions to EVC energy levels can really participate no more than ($S+1$) quantums of I-oscillations on average. On the other hand, it is known from the theory \cite{pekar1951, pekar1952, pekar1953, huang1950}, that the average of quantums of elastic oscillations of a material (phonons), participating in electron-vibrational transition on EVC is practically the same for different types of elastic oscillations and is equal $S$. Hence, maximal value $S$ for I-oscillations, for optical and acoustic phonons does not exceed ($S+1$). Taking into account that EVC are centers with $S \geq 1$, we come to a conclusion about possible values of constant $S$:
\begin{equation}
\label{eq4}
                   1 \leq S \leq S^{'}+1 = E_g /\hbar \omega_Z + 1,                \end{equation}
So, in silicon containing the A-centers ($E_g  \cong 1.1 eV$ and $\hbar\omega_Z = 0.22eV$) value $S \cong 5$, and experimental values $S$ are in limits from 1 up to 6. In InSb containing oxygen EVC  ($E_g \cong 0.2 eV$ and $\hbar\omega_Z \cong 0.22 eV$) value $S^{'} = 1$, and experimental values $S$ are in limits from 1 up to 2. Similar situation are with other semiconductors. Therefore Eq.~\eqref{eq4} can be considered as an empirical rule for estimation of value of constant $S$. Herewith it is necessary to take into account observable experimentally the reduction $S$  up to 1 and less due to elastic interaction EVC with each other at their high concentration ($N > N_{max}$).
Thus the values $T_h$ for row of semiconductors containing minimal ($N_{min}$) and maximal ($N_{max}$) concentrations EVC have been calculated under the Eq.~\eqref{eq3}. It turned out that experimental values $T_h$ for each semiconductor are between the calculated minimal and maximal hyperconductivity transition temperatures. Thus, calculated and experimental values $T_h$ coordinate with considered recombination mechanism of EVC I-oscillations generation.

The considered energy diagram of a hyperconductor is identical to the energy diagram of a superconductor in BCS theory. According to this theory, in energy spectrum of initial J-multiple degenerated (one-partial) quantum system the degeneration is removed due to paired interaction between various one-partial conditions. J-multiple degenerated discrete energy level of initial system gives rise to energy band containing (J-1) levels, and one discrete level, separated from this band on $-J\epsilon$.

Single-particle conditions in a hyperconductor are represented by I-oscillators, that is by atoms which are carrying out I-oscillations.  One can see from Fig.~\ref{fig1} that the role of an energy band in a hyperconductor is carried out by conductivity energy band (or valence energy band), and separated energy level coinciding with E*. This level separated from a conductivity band on EVC oscillation energy. Thus the role of disturbance is carried out phonons by means of which centers interact with each other, and the matrix element of this disturbance equal $-\epsilon = -(E_c - E^*)$. The corresponding energy level $(E_c - E^*)$  is shown on Fig.~\ref{fig1}. The given level has finite width due to interaction of I-oscillators with various phonons. According to measurements of phonon drag effect on EVC the average energy band width of electron-vibrational transition is equal $2 k \theta $ and $4 K < \theta < 10 K $ irrespective of temperature of material and meaning of $ \nu $ \cite{vdov2005}.
\section{Discussion}
Electronic energy spectra in superconductors and in hyperconductors have the forbidden energy gapes. Width of energy gape ($2\Delta$) in accordance with BCS theory, at temperature $T = 0 K$  and at weak electron-phonon connection, close to $3.5kT_c$. Therefore the energy gape of superconductors generally speaking, cannot be great.
However, the superconductor containing impurity atoms, whose energy levels located in the middle of energy gape, can have the big density of states and accordingly higher temperatures $T_c$. Most likely high values $T_c$ can be achieved in doped materials or in materials with complex structure and content when in energy gape of a superconductor there are energy levels with high density of electronic conditions.
One can see from the energy diagram of a hyperconductor shown on Fig.~\ref{fig1}, that  N(F) is defined by number of EVC in individual volume of a material whose I-oscillations are excited. According to the Eq.~\eqref{eq2}  the size $V^*$  can exceed  ($E_c - F$) by one quantum of I-oscillation, if a constant $S \geq 1$ . It provides values $T_c > 1490 K$ in hyperconductors.  Therefore it is obvious, that containing EVC superconductor, as well as a hyperconductor, has high densities of conditions and according to the Eq.~\eqref{eq4} $V^{*}$ becomes not less than a quantum of I-oscillations. That's why the width of gape energy  increase from $2\Delta$ up to $V^*$ and even up to the more big values. Besides, EVC increase a constant $S$ and by that promote increase $V^*N(F)$ and $T_c$. In connection with the given analogy between hyperconductors and superconductors it is possible to claim, that not only hyperconductors, but also containing EVC superconductors, metals and semimetals may have extremely high values $T_c >> T_h$. On the other hand  $T_h \approx E_{ph}/k$, where $E_{ph}$ -  energy of a phonon, participating in electron-vibrational transitions on EVC.
\begin{figure}[htbp]
\vspace*{+0.2cm}
%\mbox{\epsfxsize=3.3in \epsfbox{Fig3.eps}}
\includegraphics[width=3.3in]{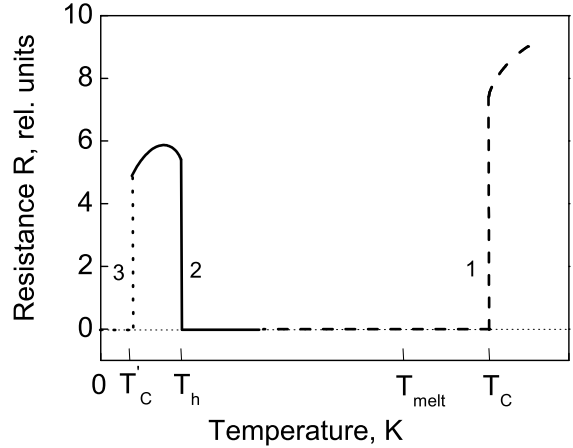}
\vspace*{-0.3cm}
\caption{Possible temperature dependence of a hyperconductor resistance R.}
\label{fig3}
\end{figure}

 The analogy between hyperconductors and superconductors also supposes probable presence of superconductivity in a hyperconductor at temperatures $T < T_h$ when I-oscillations practically are not excited. In such circumstances in a hyperconductor may be traditional superconductivity at decreasing of temperature below $T_h$.
Probable temperature dependence of  hyperconductor resistance expected from the given analogy is qualitatively represented on Fig.~\ref{fig3}.
This dependence contains a site of traditional superconductivity at temperatures between $0 K$ and $T^{'}_{c}$, a site with finite resistance at temperatures between $T^{'}_{c}$ and $T_h$, and also a site of hyperconductivity with zero electric resistance at temperatures $T_h < T < T_c$. Evidently, it is possible a situation when $T^{'}_{c}$ $ \geq $ $T^{'}_{h}$. Site of low-temperature superconductivity ($0, T^{'}_{c}$) and site of hyperconductivity ($T_h, T_c$) may adjoin with each other. Then the material has zero electric resistance at temperatures from $0 K$ up to $T_c$. In addition, on hyperconductivity site ($T_h < T < T_c$), in contrast to superconductivity site ($T < T^{'}_{c}$), for example, it is impossible to carry out the Seebeck effect because it is not  possible to create and support a gradient of temperature due to thermal superconductivity.

The physical phenomena of hyperconductivity and heat superconductivity exist due to strong electron-phonon connection provided by EVC. Electron-phonon connection  exzist on EVC due to an exchange of energy between system of atomic nuclei and electronic system of a material. In other words, these phenomena are possible only in nonadiabatic conditions, beyond the well known adiabatic Born - Oppenheimer approach \cite{born1927} and undoubtedly refer to nonadiabatic phenomena, to nonadiabatic electronics \cite{vdov2006, vdo2005}.

Hyperconductivity is put into practice experimentally in semiconductors containing electron-vibrational centers. It quite coordinate with well-known scientific representations and does not contradict physical principles. It is known, that spatial transport of  electrons (and holes) occurs mainly by means of typical for EVC electron-vibrational transitions. In such transitions (together with electrons) participate I-oscillations of atomic nuclei in atoms of a material and phonons. The zero electric resistance of a material in a wide range of temperatures is provided due to rather big energy of I-oscillations. Electrons (participating in I-oscillations) have rather big energy of I-oscillations that allows them and connected to them phonons to overcome potential barriers in materials and to create phenomena of the  hyperconductivity and thermal superconductivity. Besides, electronic transport occurs inseparably with transport of  I-oscillations and phonons, connected with electrons participating in the transitions. Transport of phonons and I-oscillations of atomic nuclei also does not meet resistance at electron-vibrational transitions, as well as at the electrons transport. Therefore, on the one hand, hyperconductivity is the superconductivity existing due to strong electron-phonon connection on EVC at temperatures between $T_h$ and $T_c$. On the other hand, hyperconductivity is inseparable from thermal superconductivity because inherent oscillations and phonons, for the same reasons, as electrons, not meets transport resistance  in a material at electron-vibrational transitions.
\section{Conclusion}
Thus, it is evident that hyperconductivity is superconductivity caused by inherent oscillations of atomic nuclei in atoms and by their connection with electrons and phonons thanks to electron-vibrational centers in materials. Hyperconductivity existence from about room temperature up to higher temperatures is proved now. Undoubtedly, hyperconductivity  is the self-supporting, independent physical phenomenon which combines superconductivity with thermal superconductivity.

Hyperconductivity phenomenon is easily realizable in practice, enables various applications and belongs to up-to-date nonadiabatic electronics.
\end{document}